\documentclass[journal=jacsat,manuscript=article]{achemso} 
\setkeys{acs}{keywords = true}
\usepackage{graphicx}
\usepackage{chemformula} 
\usepackage[T1]{fontenc} 
\usepackage{chemfig}
\usepackage{amsmath}
\usepackage{bm}
\usepackage{xcolor}

\usepackage{soul}
\usepackage[normalem]{ulem}
\newcommand{\rsmath}[1]{\bgroup\markoverwith{\textcolor{red}{\rule[0.5ex]{2pt}{0.4pt}}}\ULon {\textcolor{red}{#1}}}                                           

\author{Xiuyang Xia}
\affiliation{{School of Chemistry, Chemical Engineering and Biotechnology, Nanyang Technological University, 62 Nanyang Drive, Singapore 637459}}
\alsoaffiliation{Division of Physics and Applied Physics, School of Physical and Mathematical Sciences,Nanyang Technological University, 21 Nanyang Link, Singapore 637371}%
\altaffiliation{These authors contributed equally.}
\author{Peilin Rao}
\affiliation{{School of Chemistry, Chemical Engineering and Biotechnology, Nanyang Technological University, 62 Nanyang Drive, Singapore 637459}}
\altaffiliation{These authors contributed equally.}
\author{Juan Yang}
\affiliation{Department of Chemistry, National University of Singapore, 117546 Singapore}
\author{Massimo Pica Ciamarra}
\affiliation{Division of Physics and Applied Physics, School of Physical and Mathematical Sciences,Nanyang Technological University, 21 Nanyang Link, Singapore 637371}
\author{Ran Ni}
\affiliation{{School of Chemistry, Chemical Engineering and Biotechnology, Nanyang Technological University, 62 Nanyang Drive, Singapore 637459}}
\email{r.ni@ntu.edu.sg}

\title{Entropy driven thermo-gelling vitrimer}


\begin{document}

\begin{tocentry}
\includegraphics[width=\textwidth]{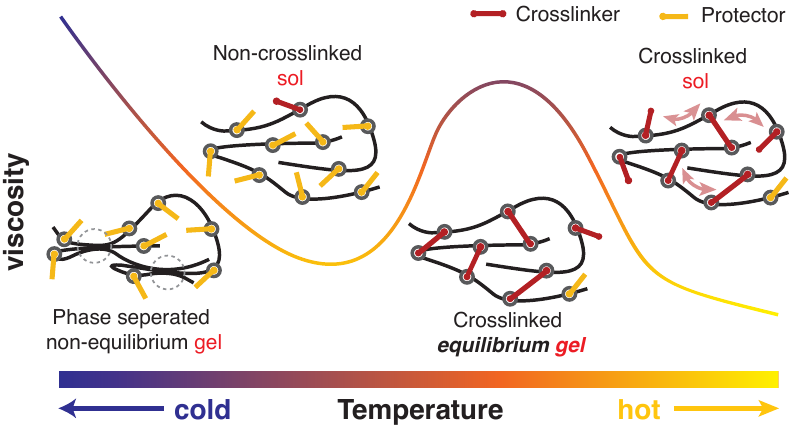}
\end{tocentry}

\begin{abstract}
Thermo-gelling polymers have been envisioned as promising smart biomaterials but limited to their weak mechanical and thermodynamic stabilities. Here we propose a new thermo-gelling vitrimer, which remains at a liquid state because of the addition of protector molecules preventing the crosslinking, and with increasing temperature, an entropy driven crosslinking occurs to induce the sol-gel transition. Moreover, we find that the activation barrier in the metathesis reaction of vitrimers plays an important role, and experimentally one can use catalysts to tune the activation barrier to drive the vitrimer to form an equilibrium gel at high temperature, which is not subject to any thermodynamic instability. We formulate a mean field theory to describe the entropy driven crosslinking of the vitrimer, which agrees quantitatively with computer simulations, and paves the way for design and fabrication of novel vitrimers for biomedical applications.
\end{abstract}

\section{Introduction}
Thermo-gels are the polymer that undergoes a sol-gel transition with increasing temperature. Over the past decades, thermo-gels have gained extensive attention in biomaterials and biomedical sciences~\cite{roy2013,alarcon2005,qiao2018,dou2014}, because of their significant potentials in applications of shape-memory polymer~\cite{andreas2002}, drug delivery~\cite{bajpai2008,nath2002,chen2008}, tissue engineering~\cite{laloyaux2010},  bioseparations~\cite{hoffman2007,Shamim2007}, etc.
The typical mechanism of thermo-gelling involves non-covalent interactions in the material such as hydrogen bonding~\cite{fujishige1989}, Coulombic~\cite{huglin1991}, and competing van der Waals interactions~\cite{bomboi2016}. 
By increasing temperature, the intra- and intermolecular interactions outbalance the solvent effect, and thereby results in the gelation of the material~\cite{kapnistos2000,zhang2008,moon2012}.
However, the weak non-covalent interaction usually leads to low mechanical stability and rapid disintegrations in physiological environments~\cite{yuan2018,dong2012}.
To overcome those challenges, one may obtain a permanently crosslinked gel networks with higher mechanical stability by using chemical reactions such as Michael addition~\cite{wetering2005}, enzymatic crosslinking~ \cite{sakai2010, lee2009}, or photo-induced crosslinking~\cite{yuan2018, vermonden2008}. However, the gel formation requires an extra re-processing step that is difficult to control or not applicable for thick or opaque samples~\cite{leeacta2011,parinaz2019,Francesco2002}, and the formed gel would lose the reversibility.
 
Vitrimers are a unique type of crosslinked polymer networks with dynamic and exchangeable covalent bonds~\cite{montarnal2011silica}.
They have shown great promises as recyclable materials in chemical and materials industries. The exchangeability originates from the reversible reactions between polymer backbones and small crosslinkers through metathesis reactions. The rate of these reactions is governed by temperature, and some of them, e.g., silyl ether metathesis~\cite{tretbar2019}, thio-disulfide exchange reaction~\cite{wu2010,pepels2013}, dynamic Schiff base~\cite{cho2018,jiang2021},
are biocompatible and applicable \emph{in vivo}.
In this work, based on a recently developed linker-mediated vitrimer using metathesis reactions~\cite{rottger2017,Lei2020}, we propose a new reversible
thermo-gelling polymer. The polymer remains as an un-crosslinked liquid at low temperature and solidifies to a covalently crosslinked elastomer with increasing temperature. To describe the crosslinking of the designed thermo-gelling  vitrimer, we formulate a mean field theory, which quantitatively agrees with coarse-grained computer simulations. Moreover, we find that the activation barrier of metathesis reactions plays a crucial role that it can drive the formation of an equilibrium gel with increasing temperature.

\section{Model and mean field theory}
\begin{figure*}[ht]
   \centering
   \includegraphics[width=1.0\textwidth]{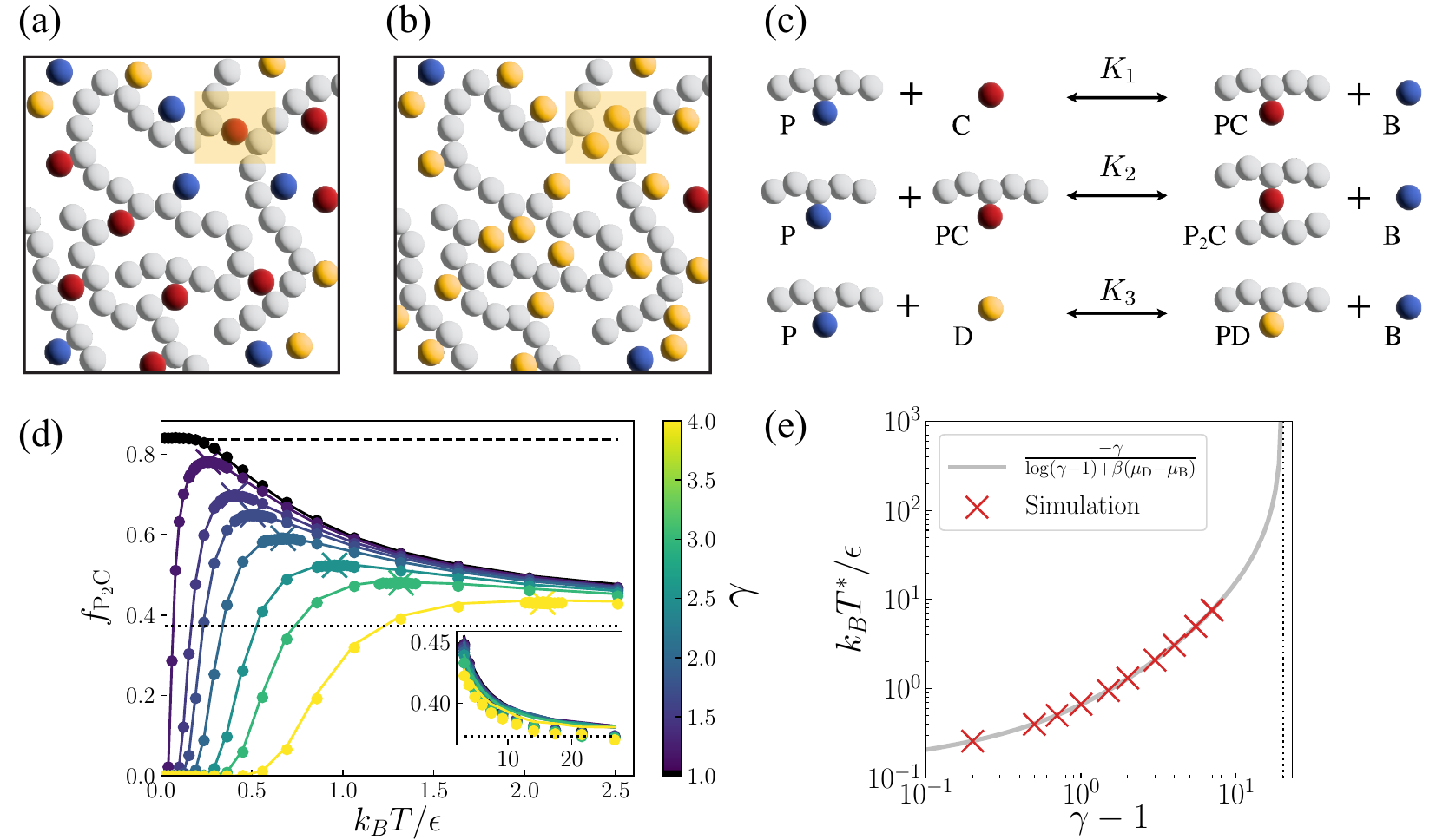}
   \caption{\label{fig1} \footnotesize \textbf{Entropy driven crosslinking vitrimers.} {(a) Schematic representation of the linker-mediated vitrimer protected by small protector molecule $\rm D$. (b) At low temperature, PD bonds dominate the system with no crosslinking formed. The rectangular yellow region indicates that a ${\rm P_2C}$ crosslinking bond transforms into two dangling $\rm PD$ bonds at low temperature. (c) Three metathesis reactions in the linker-mediated vitrimer, of which the reaction constants are $K_1$, $K_2$ and $K_3$, respectively.} (d) The crosslinking degree $f_{\rm P_2C}$ as a function of temperature $k_BT/\epsilon$ for various $\gamma$ from 1.0 to 4.0. Symbols are values obtained from simulations and solid lines are the theoretical prediction of Eq.~\ref{eq:fP2C_Gamma} using $K_2/K_1$ calculated from simulations. The dotted and dashed horizontal lines are the values of $f_{\rm P_2C}$ at the high temperature limit, and at the low temperature limit with $\gamma=1$  theoretically predicted by Eq.~\ref{eq:fP2C_Gamma} with Eqs.~\ref{eq:Gamma_inf} and \ref{eq:Gamma_0}, respectively. $\times$ symbols indicate the location of maximum value of $f_{\rm P_2C}$ obtained in simulations. Inset: $f_{\rm P_2C}$ as a function of $k_BT/\epsilon$ with at higher temperature showing the convergence to the theoretical prediction. (e) Optimal temperature $k_B T^*/\epsilon$ as a function of $\gamma$, where symbols are from simulations and the grey curve is the theoretical prediction of Eq.~\ref{eq:Ts}. The dotted vertical line shows the critical $\gamma$ where $T^*$ diverges. Here $N_{poly}=100$, $\phi_p=0.25$, $\chi_{\rm att}=0$, $\beta \mu_{\rm B} = -2$, $\beta \mu_{\rm C} = -4$ and $\beta \mu_{\rm D} = -5$.}
\end{figure*}

We consider a system of volume $V$ consisting of $N_{poly}$ polymer chains, and each polymer chain comprises $n$ connected hard spheres of diameter $\sigma$ as the backbone, on which $m$ precursors $\rm P$ are uniformly distributed.
{As shown in Fig.~\ref{fig1}a and c}, each $\rm P$ can react with a cross-linker molecule $\rm C$ or a protector molecule $\rm D $ to form a dangling $\rm PC$ or $\rm PD$ bond through metathesis reactions of reaction free energy $\Delta G_{\rm C}=-\epsilon$ or $\Delta G_{\rm D}=-\gamma\epsilon$, respectively, with releasing a byproduct molecule $\rm B$. 
Only dangling $\rm PC$ bonds can further react with another precursor $\rm P$ to form a $\rm P_2 C$ crosslinking bond through the same metathesis reaction. To ensure no crosslinking formed at the low temperature limit {(as shown in Fig.~\ref{fig1}b)}, we consider $\gamma > 1$.
Small molecules $\rm B$, $\rm C$ and $\rm D$ are modelled as hard spheres of diameter $\sigma$ and controlled by the chemical potential $\mu_{\rm B}$, $\mu_{\rm C}$ and $\mu_{\rm D}$, respectively. 
We assume that densities of small molecules in the reservoir remain constant for all temperature, i.e., $\beta\mu_{\rm B}$, $\beta\mu_{\rm C}$ and $\beta\mu_{\rm D}$ do not change with $T$. Here $\beta = 1/ k_B T$ with $k_B$ and $T$ the Boltzmann constant and temperature of the system, respectively.
The packing fraction of polymer backbones is $\phi_p=\pi N_{poly}n\sigma^3/6V$, and $N_i$ is number of bonds or molecules in the system, where $i= \rm P $, $\rm PC$, $\rm PD$, $\rm P_2C$, $\rm B$, $\rm C$ and $\rm D$.
An infinitely deep square-well potential is employed to mimic the covalent bonds in the system,
\begin{equation}
V_\mathrm{bond} (\mathbf{r},\mathbf{r}')=\left\{\begin{array}{ll}
0 & \left|\mathbf{r}-\mathbf{r}^{\prime}\right|<r_\mathrm{bond} \\
\infty & \text { else }
\end{array}\right.,
\end{equation}
where $r_\mathrm{bond}$ is the cut-off distance of the covalent bond. We use the square-well potential to model the  van der Waals short range attraction between polymer backbones:
\begin{equation}
V_\mathrm{att} (\mathbf{r},\mathbf{r}')=\left\{\begin{array}{ll}
-\chi_\mathrm{att} \epsilon & \left|\mathbf{r}-\mathbf{r}^{\prime}\right|<r_\mathrm{att} \\
0 & \text { else }
\end{array}\right.,
\end{equation}
where $\chi_\mathrm{att}$ and $r_\mathrm{att}$ are the strength and interaction range of the attraction, respectively. 
To be experimentally relevant, we choose $n=100$, $m=8$ and $r_{\rm bond} = r_\mathrm{att}=  1.5 \sigma$ throughout all simulations in this work.

The free energy of the system can be written as
 \begin{equation}\label{eq:betaF}
\begin{aligned}
{\beta F} =&   \sum\limits_{i=poly,\rm B,C,D}N_{i}{ \left[\ln \left( \frac{N_{i} \Lambda^3}{V}\right) -1\right]}
 +  \sum_{i=\rm P,PC,PD,P_2C} N_{i} \left[\ln \left( \frac{n_{i}\Lambda^3}{V_p} \right) -1\right] \\
 & -N_{\rm P_2C}k_B^{-1}  \Delta S +  \beta F_{\rm att} + \beta F^{ex}_{\rm HS}- \beta N_{\rm B}\mu_{\rm B} -\beta \left(N_{\rm PC}+ {N_{\rm P_2C}} + N_{\rm C} \right) \mu_{\rm C}
-\beta \left( N_{\rm PD} + N_{\rm D} \right) \mu_{\rm D}\\
 &+ \beta (N_{\rm PC} +2  N_{\rm P_2C} ) \left(\Delta G_{\rm C} +\mu_{\rm B} \right)
 + \beta N_{\rm PD} \left(\Delta G_{\rm D} +\mu_{\rm B} \right)
  ,
    \end{aligned}
\end{equation}   
where $\Lambda$ is the de Broglie wavelength. The first summation is on the ideal gas terms of polymer blobs ($poly$), byproduct molecules ($\rm B$), free crosslinkers ($\rm C$) and protectors ($\rm D$). The second summation is on the ideal gas terms of precursors $\rm P$, $\rm PC$ bonds, $\rm PD$ bonds and $\rm P_2C$ bonds on a polymer confined within the volume $V_p$, which can been as the volume occupied by a polymer, and $n_i=N_i/N_{poly}$  is the number of bonds per polymer with $i = \rm {P, PC, PD, P_2C}$. 
$\Delta S$ accounts for the entropy change of the system by forming a $\rm P_2C$ crosslinking bond, and $F_{\rm att}$ is the free energy contribution from the van der Waals short range attraction between polymer backbones.
It is known that the structure of a homogeneous fluid is mainly determined by the repulsion between molecules~\cite{weeks1971role}. Therefore, $F_{\rm att} \approx - \chi_\mathrm{att} \epsilon g(\phi)$, and $g(\phi)$ is a function of the total packing fraction of the system $\phi$, which can be seen as the total number of non-bonded attractive pairs of backbone beads formed in the system.
$F^{ex}_{\rm HS}$ is the excess free energy accounting for the crowding effect, which we approximate with the Carnahan–Starling hard-sphere equation of state~\cite{hansen1990theory}(see Supplementary Materials S1), and it depends on the total packing fraction of the system $\phi$. The other five terms arise from the bond swaps in metathesis reactions related with $\Delta G_{\rm C}$ and $\Delta G_{\rm D}$, and the exchange of molecules with reservoir related with $\mu_{\rm B}$, $\mu_{\rm C}$ and $\mu_{\rm D}$. 

We define the crosslinking degree of the system as $f_{\rm P_2C}=2N_{\rm P_2C}/(N_{poly}m)$ and the fraction of other bonds or molecules as $f_i = N_i / (N_{poly}m) $ with $i= \rm P, PD, PC, B, C, D$, and 
\begin{equation}\label{eq:fP2C_eq}
f_{\rm P} + f_{\rm PC} + f_{\rm PD} + f_{\rm P_2C} = 1.
\end{equation}
With the saddle point approximation, i.e., $\partial F/ \partial \{f_j\} = 0 $ with $j = \rm PC, PD, P_2C$, we have $f_{\rm PC} = f_{\rm P} \Xi_{\rm PC}$, $f_{\rm PD} = f_{\rm P} \Xi_{\rm PD}$ and $f_{\rm P_2C} = f_{\rm P}^2 \Xi_{\rm P_2C}$, where 
{
\begin{equation}
\begin{aligned}
	\Xi_{\rm P_2C} =& ( 2m\Lambda^3/V_p)\exp [{\beta(\mu_{\rm C} -2\mu_{\rm B} +2\epsilon+\mu^{ex}_{\rm  HS} )+k_B^{-1} \Delta S }]\\
	\Xi_{\rm PC} =& \exp[{\beta (\mu_{\rm C} -\mu_{\rm B} +\epsilon)}]\\
	\Xi_{\rm PD} =& \exp [{\beta (\mu_{\rm D} -\mu_{\rm B} +\gamma\epsilon )}].
\end{aligned}
\end{equation}} 
$\mu_{\mathrm{HS}}^{e x} $ is the excess free energy of hard spheres with packing fraction $\phi$, which can be calculated by self-consistent iterations (see Supplementary Materials S1).
Here we note that $\partial F_{\rm att}/\partial f_j \approx -\chi_{\rm att} \epsilon (\partial g/ \partial \phi) (\partial \phi/ \partial f_j) \approx 0 $, with the assumption that the system is mainly composed of polymer backbones, and the metathesis reactions have minor effects on the packing fraction of the system, i.e., $\partial \phi/ \partial f_j \approx 0$. Substituting these terms into Eq.~\ref{eq:fP2C_eq}, we obtain
\begin{equation}\label{eq:fP2C_Gamma}
f_{\rm P_2C} = \left( \sqrt{\Gamma^2 + 1} - \Gamma \right) ^2,
\end{equation}
with
\begin{equation}\label{eq:Gamma}
\Gamma = \frac{1+ \Xi_{\rm PC} + \Xi_{\rm PD}}{2 \sqrt{\Xi_{\rm P_2C}}}.
\end{equation}
Moreover, $V_p$ and $\Delta S$ can be obtained by considering the chemical and reaction equilibrium (see Supplementary Materials S2):
\begin{equation}
\frac{K_2}{K_1} = \frac{V}{N_{poly} V_p} e^{\Delta S/k_B},
\end{equation}
 {where $K_1$ and $K_2$ are reaction constants of the first two metathesis reactions in Fig.~\ref{fig1}c, and can be measured directly in computer simulations.}

\section{Simulations} 
\subsection{Monte Carlo simulation}
We perform Monte Carlo (MC) simulations to investigate the equilibrium properties of the thermo-crosslinking vitrimer. Assuming the bulk density conservation of $\rm B$, $\rm C$ and $\rm D$ when temperature varies, we employed grand canonical $\beta \mu_{\rm B} \mu_{\rm C} \mu_{\rm D}$--$N_{p o l y} V T$ MC simulations to simulate the coarse-grained hard-sphere-chain system, with the temperature unit $\epsilon/ k_B$. Simulations are performed with $N_{poly}$ polymer chains with $n$ beads per chain, where periodic boundary conditions are applied in all three dimensions with the cubic box volume $V=\frac{\pi}{6}N_{poly} n/\phi_p$. The translational moves of particles are implemented by the straight event-chain algorithm~\cite{bernard2009event,kampmann2015monte}, in which the length of each event is fixed at $L_c=10\sigma$, and the pressure $P$ of the system can be calculated by the mean excess chain displacement
\begin{equation}
    P=k_{B} T \rho\left\langle\frac{x_{\text {final }}-x_{\text {initial }}}{L_{c}}\right\rangle_{\text {chains }},
\end{equation}
where $x_{\mathrm{initial}}$ and $x_{\mathrm{final}}$ are the projection of initial and final particle positions on the chain direction of each event chain, $\rho$ is number density of the particles and $\langle\cdot\rangle_{\mathrm{chains }}$ calculates the average over all event chains. Molecules $\rm B$, $\rm C$ and $\rm D$ are inserted or removed using the conventional grand canonical MC method. Besides, we devise a simple bond swap algorithm to simulate reversible bond swap metathesis reactions in the vitrimer system (see below), respecting the detailed balance. The ratio of three types of trial moves, translational moves, inserting/removing molecules and bond swaps, is $1:4:5$. In each simulation, we perform $10^9$ MC moves for equilibration and $10^9$ MC moves for sampling.

\subsection{Hybrid event-driven molecular dynamics - Monte Carlo simulation} 
We perform the hybrid event-driven molecular dynamics - Monte Carlo (EDMD--MC) simulation~\cite{wu2019dynamics, stukalin2013self, hoy2009thermoreversible} to measure the intermediate scattering function $F_{\rm s}(q,t)$ and the structural relaxation time $\tau_\alpha$ in the canonical ($NVT$) ensemble starting from equilibrated configurations from our grand canonical MC simulations, {as in EDMD it is not efficient to vary the number of molecules}. All the bond swaps in metathesis reactions are done by the MC algorithm below~\cite{frenkel2001understanding}. In our simulations, $N_{poly}m$ bond swap attempts are performed per time interval $\tau_{\rm MD}$. Each simulation consists of $10^6\tau_{\rm MD}$ for equilibration and $10^6\tau_{\rm MD}$ for sampling.
{An Anderson thermostat~\cite{frenkel2001understanding,andersen1980molecular} is applied to all molecules every $\tau_{\rm ads} = 0.005\tau_{\rm MD}$ to control the temperature of the system, and we ensure that further decreasing $\tau_{\rm ads}$ does not change any simulation results.
Here $\tau_{\rm MD} = \sqrt{m_b \sigma^2/\epsilon}$ is the time unit of MD simulations with $m_b$ the mass of a polymer bead, where we assume all the small molecules, i.e., B, C and D, having the same mass as a polymer bead.}
 The self-intermediate scattering function $F_{\rm s}(q,t)$ {with wave vector $q$ and time $t$} is a common indicator for structural relaxation in MD simulations~\cite{wu2019dynamics}:
\begin{equation}
    F_{\rm s}(q,t)=\frac{1}{nN_{poly}}\left\langle \sum^{nN_{poly}}_{j=1} \exp{[i\mathbf{q}\cdot(\mathbf{r}_j(t)-\mathbf{r}_j(0))]} \right\rangle,
\end{equation}
which calculates over all backbone beads {with position $\mathbf{r}$} on the vitrimers.

\subsection{Bond swap algorithm}
In our MC simulations, a typical reversible single bond-swap reaction is
\begin{center}
\schemestart
\chemfig{\mathbf{a}_{G}(-[:45,,,,]\mathbf{b}_{B_0})(-[:315,,,,dash pattern=on 1pt off 1.5pt]\mathbf{c}_{C_0})}
\arrow{<->[\parbox{2cm}{\centering \footnotesize single\\bond swap}][]}
\chemfig{\mathbf{a}_{G}(-[:45,,,,dash pattern=on 1pt off 1.5pt]\mathbf{b}_{B_1})(-[:315,,,,]\mathbf{c}_{C_1})}
\schemestop\par
\end{center}
where \textbf{a}/\textbf{b}/\textbf{c} and $\rm G / A_0 / A_1 / B_0 / B_1$ are the specific beads/particles and types as shown in the three kinds of reactions in Fig.~\ref{fig1}c. {$\rm G$ is the type of polymer backbone beads grafted by precursors initially.} The proposed move for the forward reaction can be constructed by the following two steps:
\begin{enumerate}
  \item Choose a \emph{pivot} particle $\mathbf{a}$ (type $G$) bounded with $\mathbf{b}$ (type $B_0=$ P, PC, $\rm{P_2C}$ or PD) with probability $p(\mathbf{a})$. Note there is precisely one bond that can be swapped for each pivot particle;
  \item Choose a neighboring potentially reactive particle $\mathbf{c}$ (type $\rm C_0$ = B, C, D or PC) within the distance $r_{\rm bond}$ randomly with probability $h(\mathbf{c}_{C_0}| \mathbf{a})$;
\end{enumerate}
One can see the above bond swap move does not change the number of neighbor potentially reactive particles and bonds of $\mathbf{a}$, hence both before and after a bond swap move, $h(\mathbf{c}_{C_0}| \mathbf{a})=h(\mathbf{b}_{B_1}| \mathbf{a})$. For the first and third reaction in Fig.~\ref{fig1}c, the acceptance probability of the proposed move of forward and backward reactions are $\exp(-\beta\Delta G)$ and $\exp(\beta\Delta G)$, respectively, with the free energy of the forward reaction $\Delta G$. For the second reaction in Fig.~\ref{fig1}c, we double the forward reaction flux to equalize the flux between the two states before and after the reaction~\cite{Lei2020}.\\
Furthermore, for mimicking the actual topology and reaction dynamics of vitrimers, i.e., the covalent bond-exchange dynamics in Fig.~S3, we perform additional double bond swap reactions in the EDMD-MC simulation as follows, before and after which no chemical potential change in the system:
\begin{center}
\schemestart
\chemfig{\mathbf{a}_{G}([::-45]*4(-[:45,,,,dash pattern=on 1pt off 1.5pt]\mathbf{c}_{C_0}-[:315,,,,]\mathbf{d}_{G}-[:315,,,,dash pattern=on 1pt off 1.5pt]\mathbf{b}_{B_0}?-[:45,,,,]))}
\arrow{<->[\parbox{2cm}{\centering \footnotesize double\\bond swap}][]}
\chemfig{\mathbf{a}_{G}([::-45]*4(-[:45,,,,]\mathbf{c}_{C_1}-[:315,,,,dash pattern=on 1pt off 1.5pt]\mathbf{d}_{G}-[:315,,,,]\mathbf{b}_{B_1}-[:45,,,,dash pattern=on 1pt off 1.5pt]))}.
\schemestop\par
\end{center}
We randomly choose each specific bond swap move in three steps:
\begin{enumerate}
    \item Select a \emph{pivot} particle $\mathbf{a}$ (type $G$) bounded with $\mathbf{b}$ (type $B_0=$ P, PC, $\rm{P_2C}$ or PD) with probability $p(\mathbf{a})$;
    \item Select a neighboring potentially reactive particle $\mathbf{c}$ other than $\mathbf{b}$ (type $C_0=$ B, C, D, PB, PC, $\rm{P_2C}$ or PD) within randomly the distance $r_{\rm bond}$ with probability $h(\mathbf{c}_{C_0}| \mathbf{a})$.
    \item Select a reaction type (single-bond/double-bond) with probability $g(\mathbf{c}_{C_0}|\mathbf{b}_{B_0})$.
\end{enumerate}
Similarly, here $h(\mathbf{c}_{C_0}| \mathbf{a})=h(\mathbf{b}_{B_1}| \mathbf{a})$. $g$ accounts for the the number of possible bonds combinations, and if $\mathbf{b}_{B_0}/\mathbf{c}_{C_0}$ can undergo both single-bond swap and double-bond swap, $g(\mathbf{c}_{C_0}|\mathbf{b}_{B_0})=0.5$, otherwise, $g(\mathbf{c}_{C_0}|\mathbf{b}_{B_0})=1$. Additionally, we consider that all metathesis reactions share the same activation barrier $\Delta G_{b}$ and activation free energy is $\Delta G_{b}$+$\max{(\Delta G,0)}$. A bond swap attempt first needs to overcome the energy barrier with probability $acc_{b}=\exp{(-\Delta G_{b}/k_BT)}$, and then is accepted with the probability
\begin{equation}
    acc_{r} 
        =\min{\left[1, \frac{g(\mathbf{b}_{b_0}|\mathbf{a}_{a_0})}{g(\mathbf{a}_{a_1}|\mathbf{b}_{b_1})}\exp{(-\beta\Delta G)} \right]},
\end{equation}
where $\Delta G$ is the bond swap reaction free energy.

\section{Results}
\subsection{Entropy driven crosslinking}
First we investigate the crosslinking in the system without any short range van der Waals attraction, i.e., $\chi_{\rm att} = 0$, and in Fig.~\ref{fig1}d we plot the calculated $f_{\rm P_2C}$ from Monte Carlo (MC) simulations for various $\gamma$ as functions of temperature $k_BT /\epsilon$ in comparison with the theoretical prediction of Eq.~\ref{eq:fP2C_Gamma}, in which the only input from simulations is $K_2/K_1$. One can see that the results from computer simulation quantitatively agree with the mean field theory. At $\gamma = 1$, $f_{\rm P_2 C}$ reaches two different positive plateaus at $T \rightarrow 0$ and $T \rightarrow \infty$, respectively, with $f_{\rm P_2C} (T \rightarrow 0 ) > f_{\rm P_2C} (T \rightarrow \infty )$. When $\gamma > 1$, as shown in Fig.~\ref{fig1}d, $f_{\rm P_2C} (T \rightarrow 0 ) \rightarrow 0$, and with increasing temperature, $f_{\rm P_2C}$ increases to reach a maximum at temperature $T^*$ then decreases to approach the same plateau at high temperature, which does not depend on $\gamma$. These can be understood as follows. At the high temperature limit, $\beta \epsilon\to 0$, the crosslinking in the system is solely determined by entropy, and 
\begin{equation}
    \Gamma^\infty=\frac{1 + \Xi_{\rm PC}^\infty + \Xi_{\rm PD}^\infty}{2\sqrt{\Xi_{\rm P_2C}^\infty}}, \label{eq:Gamma_inf}
\end{equation}
with 
{
\begin{equation}
	\begin{aligned}
		\Xi_{\rm PC}^{\infty} =& \exp[{\beta\left(\mu_{\rm C} -\mu_{\rm B}\right)}]\\
		\Xi_{\rm PD}^{\infty} =& \exp[{\beta\left(\mu_{\rm D} -\mu_{\rm B}\right)}]\\
		\Xi_{\rm P_2C}^{\infty} =& ( 2m\Lambda^3/V_p)\exp [{\beta(\mu_{\rm C} -2\mu_{\rm B} +\mu^{ex}_{\rm  HS} )+k_B^{-1} \Delta S }],
	\end{aligned}
\end{equation}
}
which does not depend on $\gamma$. This can be further used to calculate $f_{\rm P_2C}^{\infty}$ using Eq.~\ref{eq:fP2C_Gamma}.
Moreover, at the low temperature limit, i.e., $\beta \epsilon \rightarrow \infty$, when $\gamma = 1$, the saddle-point solution of $\Gamma $ can be obtained by the Lagrange multiplier method (see Supplementary Materials S4):
\begin{equation}
    \Gamma^0=\frac{\Xi_{\rm PC}^\infty + \Xi_{\rm PD}^\infty}{2\sqrt{\Xi_{\rm P_2C}^\infty}}, \label{eq:Gamma_0}
\end{equation}
with which we can calculate the low temperature plateau $f_{\rm P_2C}^0$ for $\gamma = 1$ with Eq.~\ref{eq:fP2C_Gamma}. One can see that $\Gamma^\infty > \Gamma^0$, which implies $f_{\rm P_2C}^{\infty} < f_{\rm P_2C}^0$, as Eq.~\ref{eq:fP2C_Gamma} is a monotonically decreasing function of $\Gamma$.
When $\gamma > 1$, as the formation of PD bonds is more energetically favourable than that of PC bonds, the crosslinking is prevented at the low temperature limit, and at small $\gamma$, the competition between entropy driven crosslinking and enthalpy driven protection by PD bonds leads to the non-monotonic behaviour of $f_{\rm P_2C}$ with increasing $T$. The optimal temperature $T^*$, where $f_{\rm P_2C}$ reaches the maximum can be obtained by solving $\partial f_{\rm P_2C} / \partial \beta = 0 $ equivalent to $\partial \Gamma / \partial \beta =  0 $, which is
\begin{equation}\label{eq:Ts}
k_B T^* =  \frac{-\gamma\epsilon}{\log(\gamma-1)+\beta(\mu_{\rm D}-\mu_{\rm B})}.
\end{equation}
$T^*$ is a positive number when $1<\gamma<1+\exp [\beta(\mu_{\rm B}-\mu_{\rm D})]$. When $\gamma > 1+\exp [\beta(\mu_{\rm B}-\mu_{\rm D})]$, the protecting effect of PD bonds is too strong, and $f_{\rm P_2C}$ increases monotonically with increasing temperature. As shown in Fig.~\ref{fig1}e, the theoretically predicted $T^*$ agrees quantitatively with computer simulations. Intriguingly, Eq.~\ref{eq:Ts} does not need any input from simulation or experiments, and one can also see that the optimal temperature $T^*$ does not depend on the backbone packing fraction $\phi_p$ or the crosslinker chemical potential $\beta \mu_{\rm C}$,  which is confirmed in our MC simulations (Fig.~S1ac). However, the height of $f_{\rm P_2C}$ peak does depend on $\beta \mu_{\rm C}$, and it changes non-monotonically (Fig.~S1c). This non-monotonic dependence originates from an entropic effect recently found in linker-mediated vitrimers~\cite{wu2020,Lei2020,xia2020linker}. Moreover, the position and height of $f_{\rm P_2C}$ can be also tuned by changing the concentration of B and D molecules, i.e., $\beta \mu_{\rm B}$ and $\beta \mu_{\rm D}$ (Fig.~S1bd).

\begin{figure*}[ht]
   \centering
   \includegraphics[width=1.0\textwidth]{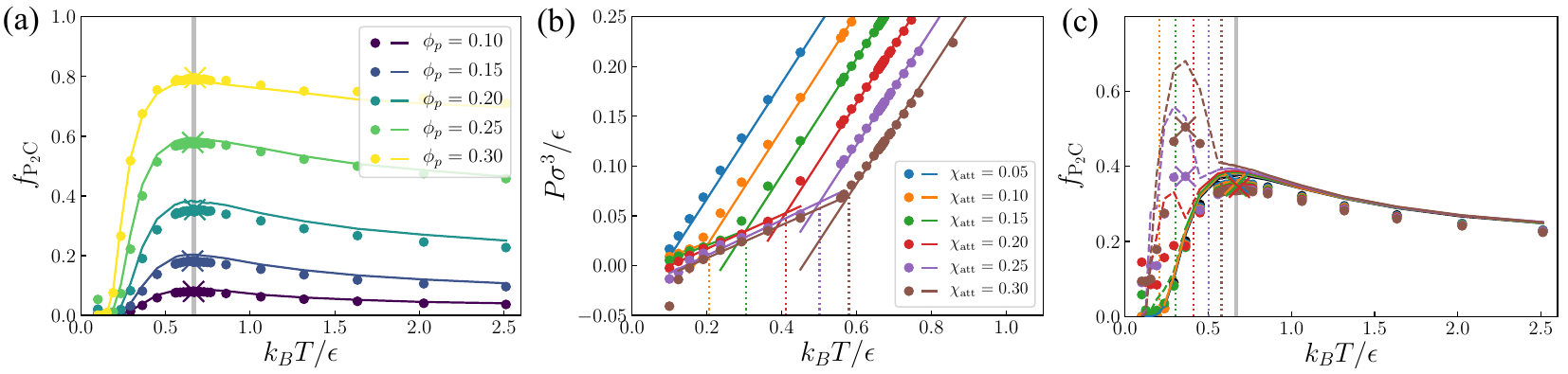}
   \caption{\label{fig2} \footnotesize \textbf{Effects of short range attraction.} (a) $f_{\rm P_2C}$ as a function of temperature $k_BT/\epsilon$ for various $\phi_p$ at $\chi_{\rm att}=0.1$. {(b,c) $f_{\rm P_2C}$ and pressure $P\sigma^3/\epsilon$ as a function of temperature $k_BT/\epsilon$ for different short range attraction $\chi_{\rm att}$. (b) The symbols are from simulations and the lines are linear fits in the two different phases, and the position of turning-points, critical temperature $T_{c}$, are indicated by dotted vertical lines. (c) Dashed and solid curves are the theoretical prediction of Eq.~\ref{eq:fP2C_Gamma} above and below the critical temperature $T_{c}$, respectively, with the measured $K_2/K_1$ from simulations. The dotted vertical lines indicates $T_{c}$ measured in (b).} Here $N_{poly}=100$, $\phi_p=0.20$, $\gamma=2.0$, $\beta \mu_{\rm B} = -2$, $\beta \mu_{\rm C} = -4$ and $\beta \mu_{\rm D} = -5$. }
\end{figure*}

\subsection{Effects of short range attraction between polymers} 
\begin{figure*}[ht]
   \centering
   \includegraphics[width=1.0\textwidth]{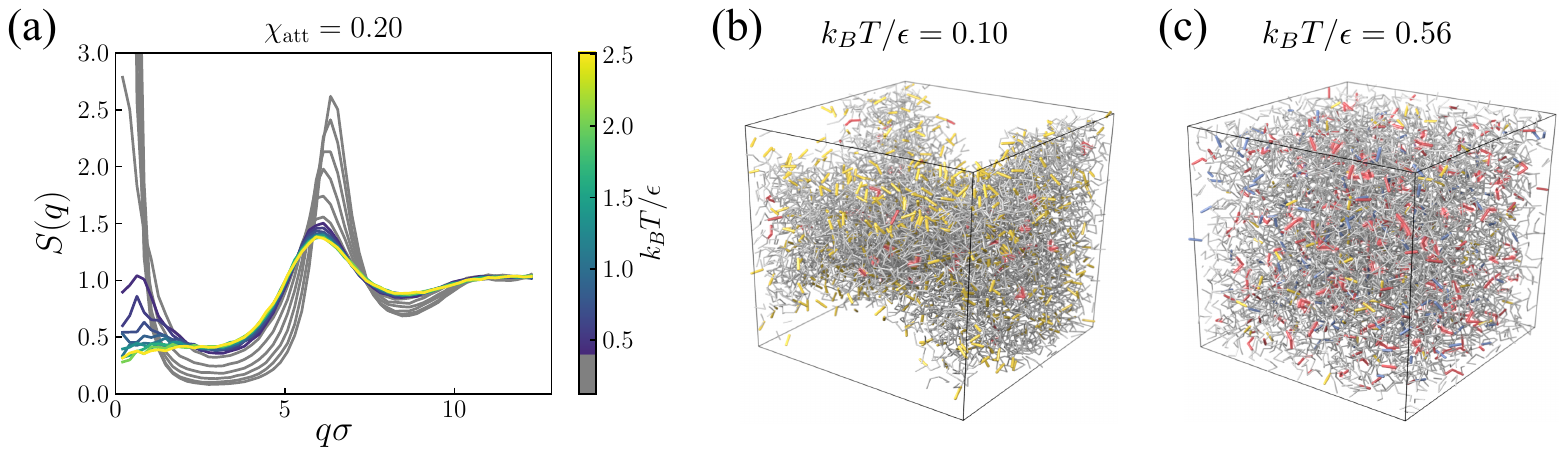}
   \caption{\label{fig3} \footnotesize \textbf{Structural change with decreasing temperature.} (a) Structure factor $S(q)$ for systems at different temperature, where the curves below $T_{\rm c}$ are grey.  {(b,c) Simulation snapshots for systems with (b) $k_BT/\epsilon = 0.10$ and (c) $k_BT/\epsilon = 0.56$. Different colors represent different types of bonds based on Fig.~\ref{fig1}c, i.e., $\rm P$, $\rm PC/P_2C$, $\rm PD$ and polymer backbone bonds are in blue, red, yellow, and gray, respectively.}   Here $N_{poly}=100$, $\phi_p=0.20$, $\gamma=2.0$, $\beta \mu_{\rm B} = -2$, $\beta \mu_{\rm C} = -4$ and $\beta \mu_{\rm D} = -5$. }
\end{figure*}

As in realistic vitrimers, there are always some short range van der Waals attraction between polymer backbones driving the gelation at low temperature, we investigate the system with weak attraction. In Fig.~\ref{fig2}a, we plot $f_{\rm P_2C}$ as a function of $k_BT/\epsilon$ for systems at $\chi_{\rm att} = 0.1$ of various polymer packing fraction $\phi_p = 0.1$ to 0.3 and $\gamma=2$. One can see that similar to the system without attraction, with increasing temperature, $f_{\rm P_2C}$ first increases and then decreases to approach a plateau at high temperature, and the results from computer simulation agree quantitatively with the theoretical prediction (Eq.~\ref{eq:fP2C_Gamma}). We further simulate the vitrimer system with changing $\chi_{\rm att}$ at $\phi_p = 0.2$ and $\gamma=2$, and the results are shown in Fig.~\ref{fig2}c. At high temperature, the simulation results agree very well with theoretical prediction, while below certain temperature $T_c$, the discrepancy appears. 
To understand the cause for the discrepancy, we plot the corresponding pressure $P \sigma^2/\epsilon$ as a function of temperature in the upper panel of Fig.~\ref{fig2}b. We see that with decreasing temperature, the slope of $P-T$ curve has a turning point coinciding with the temperature threshold $T_c$ (as indicated by the dotted vertical lines), below which the difference between simulation and theory appears.
As the slope change in $P-T$ curve normally suggests a phase separation, we plot the structure factor of the system with $\chi_{\rm att}$ at various temperature in Fig.~\ref{fig3}a and Fig.~S2. One can see that below $T_c$, $S(q)$ diverges at small $q$, which is a signature of phase separation. {This can be also seen in the snapshots of the system at different temperature, and as shown in Fig.~\ref{fig3}b and c, when temperature drops below $0.5$ $ \epsilon /k_B$, e.g., $k_B T/ \epsilon = 0.1 $ in Fig.~\ref{fig3}b,   the short range attraction induced phase separation occurs creating interfaces, which is different from the homogeneous structure at high temperature, e.g., $k_B T/ \epsilon = 0.56 $ in Fig.~\ref{fig3}c.} This suggests that the discrepancy between the theoretical prediction and computer simulations at low temperature is due to the attraction induced phase separation, which creates interfaces in the system. Although in the thermodynamic limit, the interfacial free energy is negligible, in practice, polymer systems can hardly reach a fully phase separated equilibrium state, and this makes the free energy in Eq.~\ref{eq:betaF} not able to correctly model the system.

\begin{figure*}[ht]
   \centering
   \includegraphics[width=1.0\textwidth]{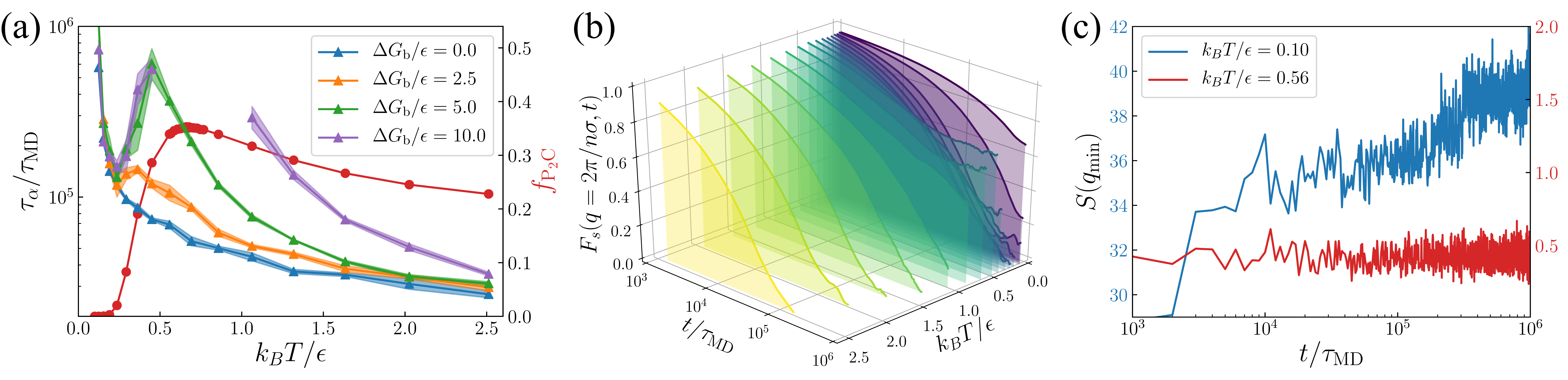}
   \caption{\label{fig4} \footnotesize \textbf{Thermo-gelling vitrimers.} (a) The structural relaxation time $\tau_\alpha$ (left) and crosslinking degree $f_{\rm P_2C}$ (right) as a function temperature $k_BT/\epsilon$ for various activation barrier $\Delta G_\mathrm{b}$. We note that $f_{\rm P_2C}$ does not depend on $\Delta G_\mathrm{b}$. (b) The intermediate scattering function $F_{\rm s}(q,t)$ with $q=2\pi / n\sigma$ at various temperature $k_BT/\epsilon$ with $\Delta G_\mathrm{b}/\epsilon=10.0$. (c) Structure factor $S(q_{\min})$ with $q_{\min} = 2\pi/V^{1/3}$ as a function of $t$ at different temperature.  Here $\phi_p=0.20$, $\gamma=2.0$ and $\chi_\mathrm{att}=0.1$ with $N_{\rm B}$, $N_{\rm C}$ and $N_{\rm D}$ obtained from MC simulations with $\beta\mu_\mathrm{B}=-2$, $\beta\mu_\mathrm{C}=-4$ and $\beta\mu_\mathrm{D}=-5$.}
\end{figure*}
 
\subsection{Thermo-gelling vitrimers} 
Next, we perform molecular dynamics (MD) simulations to investigate the thermo-induced gelation of vitrimers. We assume all metathesis reactions sharing the same activation barrier $\Delta G_{\mathrm{b}}$, and the rate of reaction $k = \nu_0 \exp \left(-\beta[\Delta G_{\mathrm{b}}+\max{(\Delta G,0)]} \right)$ with $\Delta G$ the reaction free energy. $\nu_0=1/\tau_{\rm MD}$ is the kinetic prefactor. The activation barrier $\Delta G_{\mathrm{b}}$ does not change any thermodynamic property of the system, while it does control the dynamic properties~\cite{wu2019dynamics}. We perform hybrid event-driven molecular dynamics - Monte Carlo (EDMD-MC) $NVT$ simulations for a system with $\phi_p =0.2$, $\gamma = 2.0$ and $\chi_{\rm att} = 0.1$ at different temperature starting from equilibrated configurations from grand canonical MC simulations. We plot the structural relaxation time $\tau_{\alpha}$ as a function of temperature in comparison with the corresponding $f_{\rm P_2C}$ in Fig.~\ref{fig4}a, which is defined as $F_s(q,\tau_{\alpha}) = e^{-1}$ with $F_s(\cdot)$ the intermediate scattering function. As here the diffusion is mainly through the reptation of polymer chains, we choose $q=2\pi/(n\sigma)$. One can see that when $\Delta G_{\rm b} = 0$, $\tau_{\alpha}$ decreases monotonically with increasing temperature, despite the increased crosslinking degree $f_{\rm P_2C}$. This is because that increasing temperature increases the mobility of molecules in the system, and due to the zero activation barrier of metathesis reaction, the crosslinking has minimal effects on the dynamics of the system. This is similar to the first generation vitrimer, 
which is a liquid at high temperature although fully crosslinked~\cite{montarnal2011silica}. With increasing $\Delta G_{\rm b}$, the effect of crosslinking on the dynamics of the system becomes more pronounced, and $\tau_{\alpha}$ starts to increase with temperature at about $0.3\epsilon/k_B$ to develop a peak, where $f_{\rm P_2C}$ starts to increase as shown in Fig.~\ref{fig4}a. When $\Delta G_{\rm b} =10\epsilon$, 
$\tau_{\alpha}$ diverges between $k_B T/\epsilon = 0.5$ and 1.0, 
and the corresponding $F_s(q,t)$ are shown in Fig.~\ref{fig4}b, 
which develops a plateau at both $k_BT/\epsilon \le 0.1$ and $0.5<k_BT/\epsilon < 1$. This implies that the system becomes a gel at both low and high temperature. To distinguish the property of gels formed at different temperature, we plot the time evolution of $S(q_{\min})$ in Fig.~\ref{fig4}c, which shows the large-scale structural change in the system. One can see that at the low temperature, i.e., $k_BT/\epsilon = 0.1$, $S(q_{\min})$ increases significantly with time suggesting the system undergoing a kinetically arrested phase separation {with very small amount of $\rm PC/P_2C$ bonds (Fig.~\ref{fig3}b)}, which is a typical signature of gelation induced by short range attraction~\citep{Lu2008}. However, as shown in Fig.~\ref{fig4}c, in a high temperature gel, i.e., 
$k_BT/\epsilon = 0.56$, $S(q_{\min})$ does not change with time, which suggests that there is no structural change in the system, and the thermo-gelling vitrimer is an equilibrium homogeneous gel (Fig.~\ref{fig3}c) not subject to any thermodynamic instability~\cite{sciortino2017equilibrium,Kob2007}.

\section{Discussion and Conclusion}
In conclusion, we have proposed a novel thermo-gelling vitrimer, in which at low temperature, the crosslinking is prevented by the protector, i.e., D molecules, and with increasing temperature, the entropy driven crosslinking occurs, which drives the gelation of the system. We formulate a mean field theory to describe the crosslinking, which agrees quantitatively with coarse grained computer simulations above the gas-liquid phase separation temperature induced by the short range van der Waals attraction in the system. Further molecular dynamics simulations show that the activation barrier of the metathesis reaction plays a crucial role during the thermo-gelling process, and it can be controlled by using different catalysts experimentally. Moreover, we show that the resulting thermo-induced gelation is an equilibrium gel not subject to any thermodynamic instability. 
Similar thermo-gelling systems were studied previously in patchy colloidal systems~\cite{Roldan-Vargas2013} and DNA nano stars~\cite{bomboi2016}, {and the effect was also exploited in other soft matter systems involving valency by adding corresponding protector molecules, e.g., the DNA or metal coordination-mediated nanocrystals~\cite{xia2020linker,kang2022colorimetric} and polyacrylamide-based hydrogels~\cite{fitzsimons2021effect}.}
 The difference is that the designed vitrimers here are covalently crosslinked, which is essentially a thermo-crosslinking elastomer. 
Because of the nature of metathesis reactions in vitrimers, the activation barrier can be tuned using catalysts to realize the equilibrium gel with increasing temperature, and this is qualitatively different from conventional gelations induced by short ranged attraction, which are essentially kinetically arrested phase separations~\cite{Lu2008}. 
Moreover, at high temperature, different from the thermo-gelling systems using weak interactions, e.g., hydrogen bonds, our designed vitrimers are always crosslinked because of entropy, and mechanical properties can be tuned reversibly \emph{in situ} by changing the activation barrier using catalysts and/or adjusting the concentration of B and D molecules, which is the novelty of our designed vitrimers. 
The concept of vitrimers is a versatile tool that is applicable to nearly all types of polymers. Therefore, by adopting an appropriate polymer system, the thermo-gelling vitrimer designed here should be promising for
biomedical applications.
For example, based on the designed scheme, one could employ biocompatible silyl ether metathesis to fabricate thermo-responsive biomaterials for tissue repair and injectable implants in non-invasive surgeries~\cite{tretbar2019,parrott2010}.

\begin{acknowledgement}
We thank Dr. Shilong Wu and Prof. Quan Chen in Chinese Academy of Sciences for fruitful discussions. This work is supported by the Academic Research Fund from Singapore Ministry of Education Tier 1 Gant (RG59/21) and Tier 2 Grant (MOE2019-T2-2-010). \textbf{Author contributions:} R.N. conceived the research; X.X. formulated the mean field theory and performed the Monte Carlo simulations; P. R. performed the molecular dynamics simulations; all authors discussed the results and wrote the manuscript. \textbf{Competing interests:} The authors declare that they have no competing interests. \textbf{Data and materials availability:} All data needed to evaluate the conclusions in the paper are presented in the paper and/or the Supplementary Materials. Additional data related to this paper may be requested from the authors.
\end{acknowledgement}

\begin{suppinfo}
{
\begin{itemize}
    \item SI.pdf: Additional theory details and figures including $f_{\rm P_2C}$ and the structure factor with various parameters and the description on additional double bond-swap reaction schemes. 
\end{itemize}
}



\end{suppinfo}

\bibliography{ref}

\end{document}